# Efficient Machine Learning Approach for Yield Prediction in Chemical Reactions

Supratim Ghosh,[a] Nupur Jain,[a] and Raghavan B. Sunoj [a,b,*]

[a] Department of Chemistry, Indian Institute of Technology Bombay, Powai, Mumbai 400076, India.

[b] Center for Machine Intelligence and Data Science, Indian Institute of Technology Bombay, Powai, Mumbai 400076, India.

**Abstract**

Developing machine learning (ML) models for yield prediction of chemical reactions has emerged as an important use case scenario in very recent years. In this space, reaction datasets present a range of challenges mostly stemming from imbalance and sparsity. Herein, we consider chemical language representations for reactions to tap into the potential of natural language processing models such as the ULMFiT (Universal Language Model Fine Tuning) for yield prediction, which is customized to work across such distribution settings. We contribute a new reaction dataset with more than 860 manually curated reactions collected from literature spanning over a decade, belonging to a family of catalytic *meta*-C(sp$^2$)−H bond activation reactions of high contemporary importance. Taking cognizance of the dataset size, skewness toward the higher yields, and the sparse distribution characteristics, we developed a new (i) time- and resource-efficient pre-training strategy for downstream transfer learning, and (ii) the CFR (classification followed by regression) model that offers state-of-the-art yield predictions, surpassing conventional direct regression (DR) approaches. Instead of the prevailing pre-training practice of using a large number of unlabeled molecules (1.4 million) from the ChEMBL dataset, we first created a pre-training dataset SSP1 (0.11 million), by using a substructure based mining from the PubChem database, which is found to be equally effective and more time-efficient in offering enhanced performance. The CFR model with the ULMFiT-SSP1 regressor achieved an impressive RMSE of 8.40±0.12 for the CFR-major and 6.48±0.29 for the CFR-minor class in yield prediction on the title reaction, with a class boundary of yield at 53 %. Furthermore, the CFR model is highly generalizable as evidenced by the significant improvement over the previous benchmark reaction datasets.

**Introduction**

The work embodied in this manuscript is at the interface of chemical catalysis and machine learning (ML). The introduction is therefore intended to provide a balanced overview of the topic of our investigation, first focusing on the importance of our problem selection, current status of molecular ML as applied to chemical catalysis, and the practical issues with its implementation to a rather complex situation such as a chemical reaction that involves multiple reacting components. These three aspects are briefly touched upon in the next few paragraphs, before we set forth the key motivation and objectives of this work.

The site selective functionalization of arenes could become challenging owing to the general inertness of C–H bonds in such compounds as well as their omnipresence in organic molecules. Transition metal-catalyzed C–H bond activation, facilitated by directing groups (DG) has emerged as an effective method for selective functionalization of arenes. This approach enables step- and time-economic routes for the synthesis of diverse molecular frameworks (Figure 1a).[1] The rapid advancements in the domain of C–H bond activation reactions have made it a valuable protocol in a wide range of applications such as in natural product synthesis, pharmaceuticals, organic materials, agrochemicals, polymers, dyes and so on (Figure 1b).[2]

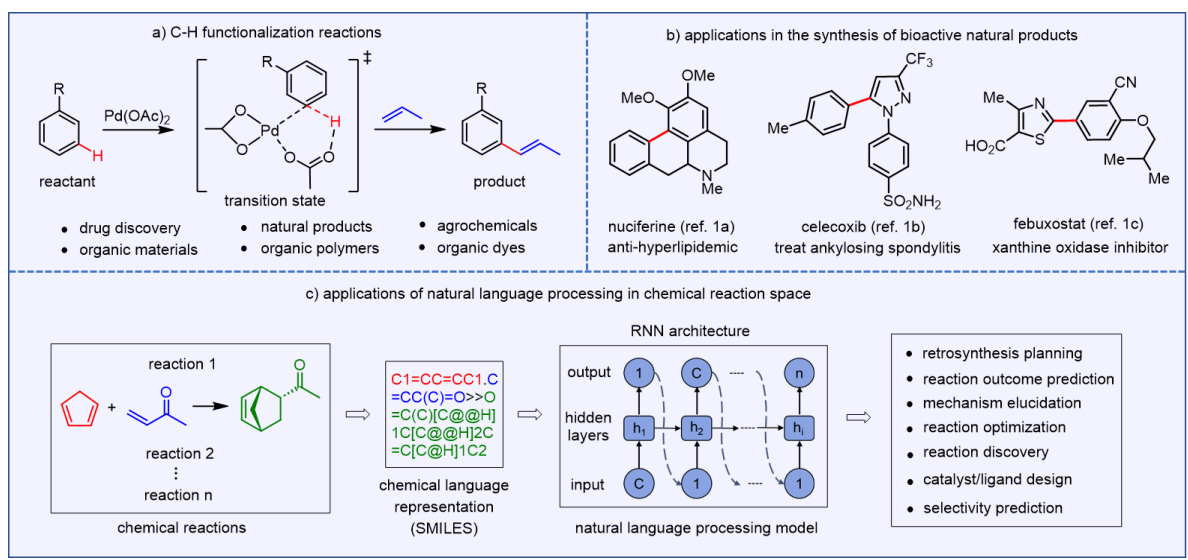

**Figure 1.** a) A general scheme representing C–H functionalization reactions, and b) its application in the synthesis of biologically active molecules. c) Natural language processing (NLP)-based deep learning framework used in studying important aspects of chemical reactions.

Gaining importance of the template or the directing group (DG) based strategies in distal C–H bond functionalization of arenes became more conspicuous in very recent years.[3] An "U-shaped" template containing a weakly coordinating nitrile DG (Figure 2a) by the Yu group[4] set the stage for ensuing developments, ably complemented by the pioneering contributions by Maiti, Yu, Li, Tan, and others.[5] Careful perusal of reaction optimization and substrate scope exploration, as documented in a series of papers, reveal that chemical intuition and mechanistic hypothesis could help make qualitative connections to the observed reaction outcome.[6] Wealth of literature in this front suggests that a trial and error approach is inevitable toward obtaining a reasonably optimal yield in these reactions.[7] A practitioner is intuitively aware that subtle changes to the key components, such as the nature of the catalyst, reaction conditions, substrates etc., often exert a pivotal influence on the reaction outcome.[8] It is therefore of high timely significance to ask whether predictive models, akin to various QSAR or LFER approaches of bygone years,[9] could be envisaged that work in tandem to assist development of new catalytic reactions.

The immanent limitations of the conventional heuristic approaches, serve as a motivation toward developing faster and sustainable reaction discovery workflows that place lower demands on time, material, and human resources.[10] One such promising approach is to employ ML, particularly the ones built on datasets derived from a relatively smaller number of known reactions.[11] The ML algorithms are inherently poised to deal with high-dimensional problems wherein the output typically shows a complex dependency on a set of input parameters, which are often convoluted. The catalytic C–H functionalization reactions, as described in Figure 1a, might therefore become a research problem that could benefit through ML intervention.[12] Application of ML algorithms on such reaction datasets would help discern the intricate patterns in the data and contribute toward making informed decisions for optimizing the reaction yields in the form of identifying better substrates/condition/catalysts, etc.[13] The ML approaches can as well supplement the intuition-based methods owing to its predictive capabilities on chemical reaction datasets.[14]

While ML has already made significant inroads into drug discovery,[15] synthesis planning,[16] catalyst design,[17] and molecular generation,[18] its deployments to reaction yield prediction tasks continue to present challenges of varying kind.[19] Among the available molecular ML models,[20] predictive modeling using language-like representations of molecules such as by using the SMILES (simplified molecular-input line-entry system)[21] encoding seems to be relatively better for deep learning architectures such as an RNN[22] and/or transformers (Figure 1c).[23] Some of the prominent language models that found excellent applications in this space are ULMFiT,[24] BERT,[25] T5-chem,[26] FP-BERT,[27] BARTSmiles,[28] and ChemBERTa.[29]

One of the known obstacles in the design of robust ML models for reaction outcome prediction stems from the scarcity of labeled data.[30] It is seen that the available chemical space, which could have been accessed through reactions between all the compatible reactants, remain only sparsely populated. Interestingly, some high throughput experiments (HTE) offer a fuller range of combinations offered by a handful of reactants.[31,32] The use of transfer learning (TL) is suggested as an effective alternative in the low-data regimes.[33] Typically, a deep learning model is first pre-trained on a large number of molecules collected from a chemical database (e.g., ChEMBL and Zinc)[34] in a self-supervised manner, which is then followed by fine-tuning on a target task of chemical space of immediate interest.[35] While these popular datasets contain millions of unlabeled biologically important molecules, there is zilch of details about chemical reactions in there. The USPTO dataset,[36] on the other hand, encompasses a wide array of chemical reactions, although biased towards successful reactions. Such distributions of labels/yields call into question their suitability for target aware pre-training tasks desirable for general-purpose applications to chemical reactions.

Furthermore, the pre-training on large unlabeled molecular datasets in a TL setting might demand rigorous hyperparameter tuning and task specific optimization, rendering them computationally expensive and time consuming. Although pre-training on large and diverse kinds of datasets might offer improved adaptability, their weights and biases focusing on specific tasks might even hinder effective knowledge transfer to new tasks, affecting the model performance and generalization capabilities. It would therefore be of interest to evaluate smaller chemical libraries bearing information pertinent to the target task, such as chemical reactions, for optimal exploitation of TL capabilities.[37] In the present context, we acknowledge that accessible datasets with real-world chemical reactions generally exhibit class imbalance, sparse distribution, besides having varying noise levels.[38] Thus, the TL-based ML models designed for yield prediction tasks should remain cognizant of the above-mentioned aspects.[39]

In view of the prospects and challenges in the effective deployment of TL-based ML models for yield prediction, the high contemporary importance of *meta*-C(sp$^2$)–H bond activation reactions, and the lack of robust ML models for this reaction class, we became interested in i) contributing a comprehensive literature-mined open access manually curated dataset for the title reaction, ii) deciphering the distribution characteristics of the dataset, iii) developing a novel pre-training protocol for reaction specific applications, and iv) building a TL-based chemical language model for the yield prediction. Herein, we propose a new strategy, named as CFR (classification followed by regression) for improved yield predictions that take into consideration some of the inherent distribution issues typical of reaction datasets.

**Methods**

**Reaction dataset:** The *meta*-C(sp$^2$)–H bond activation reaction dataset, henceforth referred to as m-CHA, is manually collected from 26 peer reviewed articles published by the Maiti, Yu, Li, Tan, Jin, and Zhou groups.[40] In particular, the curated dataset consists of transition metal catalyzed *meta*-C(sp$^2$)–H bond activation reactions facilitated by the nitrile directing group (DG) attached to an aryl moiety through a linker group (Figure 2a). The dataset contains 866 reactions that differ in terms of one or more species involved (e.g., substrate, coupling partner, catalyst, ligand, oxidant, base, and solvent). A typical sample in our dataset is a reaction, comprising of a combination of these species and an associated output value expressed in terms of the corresponding % yield that ranges from 0 to 100. The problem of interest is therefore a regression task over a labeled chemical reaction dataset.

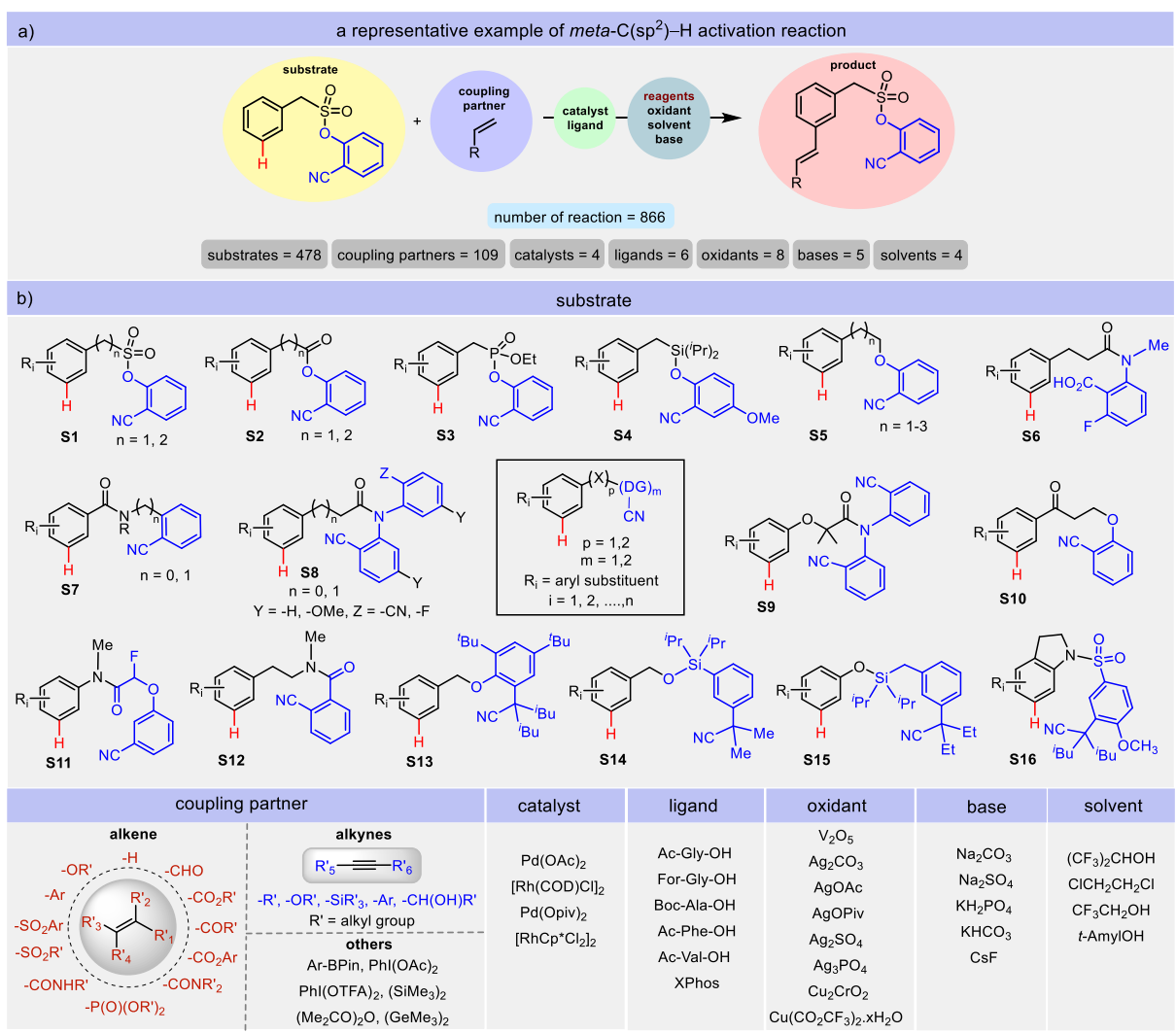

**Figure 2.** a) A generalized representation of *meta*-C(sp$^2$)–H bond activation reaction (abbreviated as m-CHA). b) Details of the substrates, coupling partners, and other species involved in the reaction.

The *meta*-C(sp$^2$)–H bond activation reaction is a widely recognized method for functionalizing diverse range of arenes. Our m-CHA dataset spans over 16 different classes that differ in the nature of the linker group (X) as well as the directing group (DG) attached to the aryl moiety (Figure 2). With this kind of diversity among the substrates, our reaction space consists of 478 arenes. Similarly, a total of 109 unique coupling partners used in functionalization can give rise to products bearing alkenes, alkynes, aryl boronates etc., on the aryl substrate. The reaction space extends further due to the use of different transition metal catalysts, ligands, bases, oxidants, and solvents. Since all these reaction variables have distinctive roles in the mechanism of the reaction, their adequate representation in the dataset is vital to a meaningful featurization for ML model building.

With the rich and diverse reaction space in the m-CHA reaction, we now focus on their molecular representation to make it conducive for ML model building. Inspired by the analogy between a string-based linguistic representation such as SMILES for molecular notation[41] and natural language processing (NLP) models, the individual reaction variables are concatenated together, separated by a dot, to provide a composite representation of the chemical reactions. In this study, a sample is a reaction that consists of concatenated SMILES of its individual molecules participating in the reaction.

**Overview of the ML model:** The ULMFit (Universal language model fine tuning) is a transfer learning-based language model (LM) originally developed for NLP tasks, wherein a sequence of words are analyzed, so as to learn to predict the next word with the highest probability in a self-supervised manner.[42] We trained the ULMFiT for molecular task following two key steps; i) the LM is first trained on a large library of unlabeled molecules, represented in the form of the corresponding SMILES string, using a multi-layer LSTM based framework. The training allows the LM to capture the extensive and in-depth knowledge of molecular language, ii) the language representation thus acquired is used in fine-tuning on a smaller set of labeled data for the intended downstream classification/regression tasks. A schematic representation of the ULMFiT transfer learning model is shown in Figure 3.

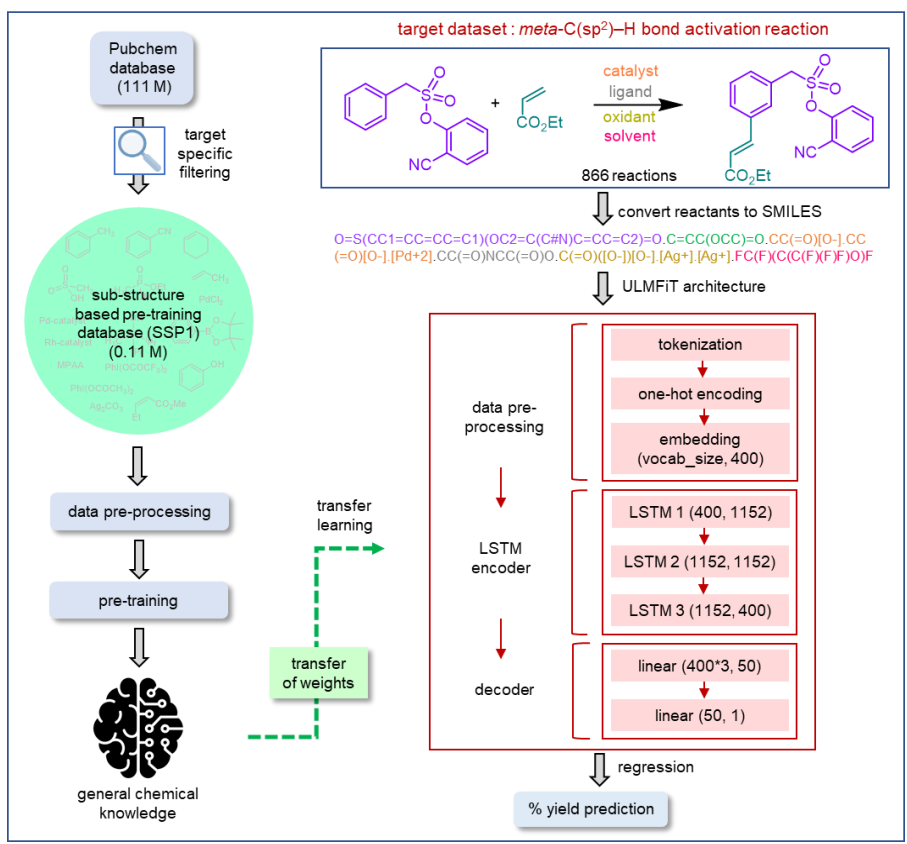

**Figure 3.** A general overview of the transfer learning model (shown on the left side) and the ULMFiT architecture consisting of an LSTM based encoder showing the number of neurons in each layer (shown in the inset on the right side).

**Results and Discussion**

We have organized our major findings into three sections, in the order of increasing importance to the overall workflow. First, an analysis of the sparsity in the reaction dataset is in focus, followed by the details of our pre-training strategy. Next, efficient TL techniques are proposed for the yield prediction on the title reaction, highlighting our novel CFR (classification followed by regression) model and our efforts to enhance both model interpretability and generalizability. In the last section, evaluation of model performance on high throughput experimentation (HTE) dataset and its comparison to the m-CHA dataset is provided, besides extension to other datasets and benchmarks.

**Sparsity in the dataset:** It is important to develop some broad understanding of the inherent characteristics present in a given dataset. The reaction space of the m-CHA is therefore analyzed using a heatmap to identify how many times each reaction component partners with every other species in the actual reaction as previously reported. A few readouts from the heatmap as given in Figure 4a are; i) the substrate-coupling partner combinations are visibly skewed toward a handful of reaction types, with the olefination reaction (594 out of 866 reactions) as the most occurring one, and ii) a high frequency of occurrence among the catalysts, ligands, and oxidants is respectively due to Pd(OAc)$_2$, Ac-Gly-OH, and Ag$_2$CO$_3$, with respect to the key substrate arene (denoted as S1, S2, …, and S16) that undergoes the C–H functionalization reaction. These characteristics in the dataset suggest that the reported experimental exploration of the chemical space, what could have ideally been a much more vast, remains just sub-optimal, if not grossly under exploited. Such a sparse distribution is expected to make the ML model building a relatively harder pursuit (*vide infra*). However, such are the datasets one would encounter in real-world reaction development, where the initial goal is to demonstrate the applicability of a newly developed reaction (e.g., *meta*-C(sp$^2$)–H bond activation reaction) across an array of substrates and coupling partners. For ML to become a valuable tool in reaction development, it should be set to work efficiently for this kind of sparse and imbalanced reaction dataset.

The challenges in ML model building for sparse datasets can well be appreciated on the basis of performance comparison of same ML model on certain HTE datasets. The inherent differences in chemical diversity and yield distribution between the m-CHA and other HTE datasets such as Suzuki coupling (SC)[31] and Buchwald-Hartwig (BH)[32] reactions are therefore worth considering at this point. Figure 4b provides a quick estimate of the theoretically accessible chemical space that considers the combinatorial possibilities between the substrate, coupling partner, catalyst, ligand, oxidant, base, and solvent, leading to a fuller set of reactions.

While the HTE datasets considered here exhibit a denser coverage within its limited number of reaction partners as employed (< 20K reactions), the m-CHA dataset contains very few reactions from among its accessible range of well over 10 million reactions. With only about 866 reported reactions, our m-CHA dataset is obviously much more sparser than the HTE counterparts. It is important to note that in the current practice, most of the new ML models for chemical reaction outcome prediction are benchmarked against such HTE datasets.[38b,24a-b,43] Another distribution aspect is that these HTE datasets also exhibit a relatively more homogeneous distribution of the yield as compared to the m-CHA dataset, where the yield values are skewed toward the higher end.[44]

a) Heatmap analysis: exploring relative occurrence and diversity in the reported experimental data (see Figure 2 for the identities of S1, S2,...,S16)

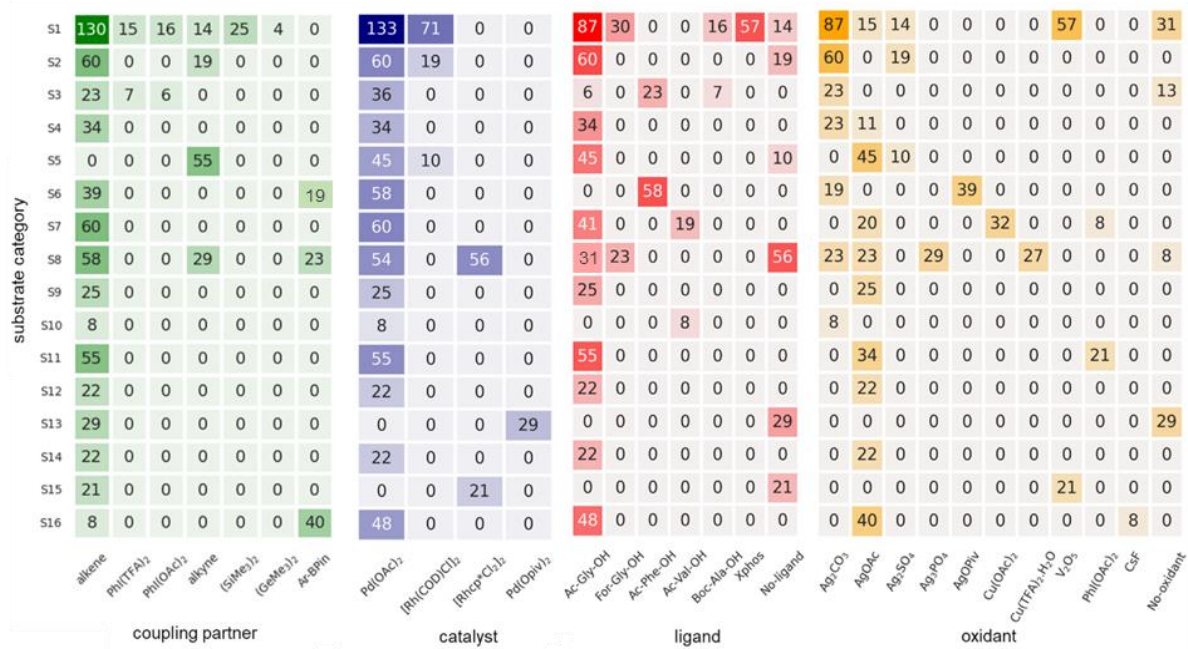

b) HTE versus m-CHA dataset

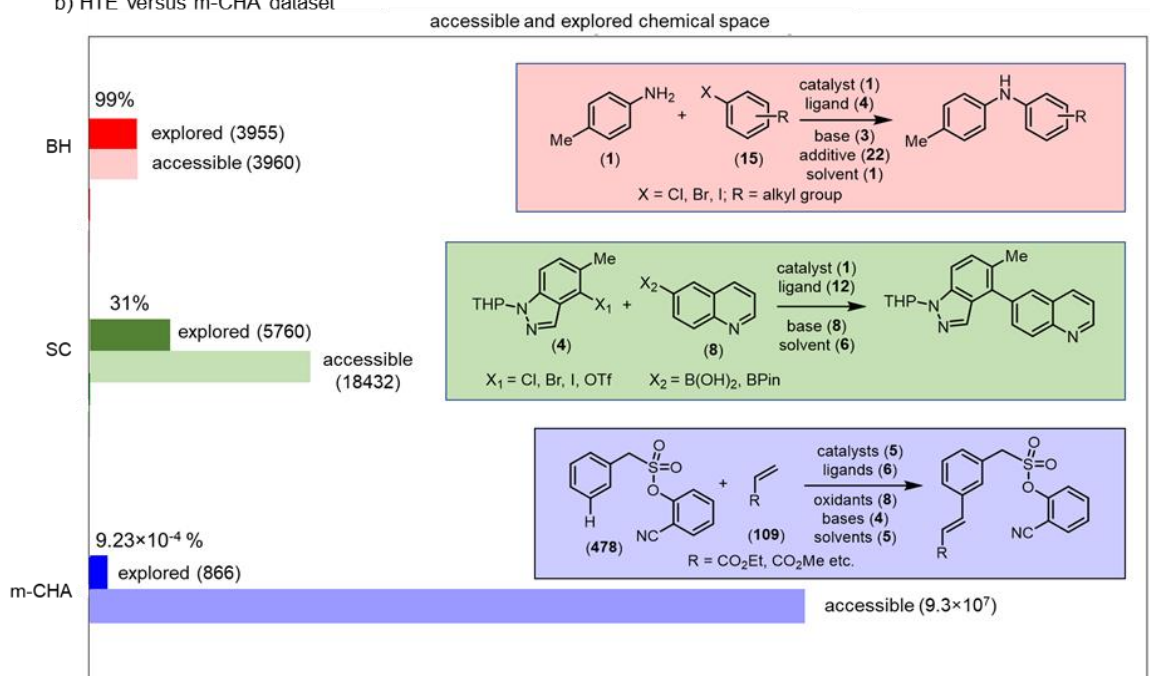

c) Visualization of chemical reaction space

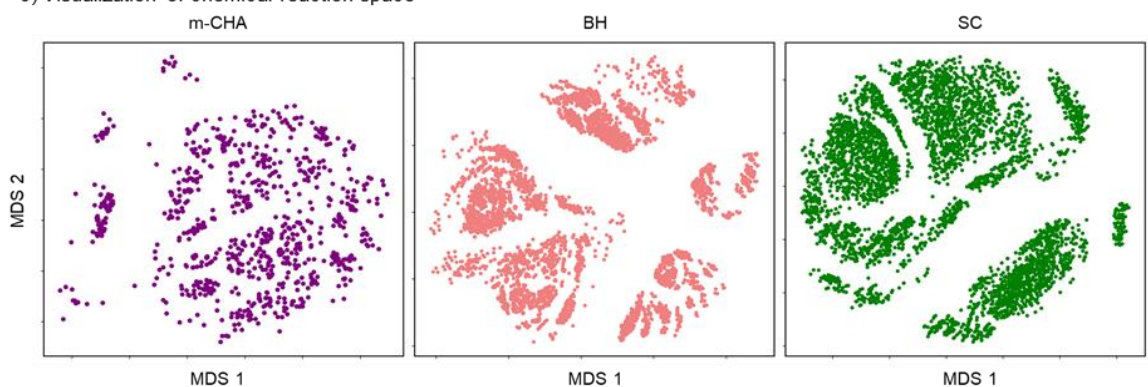

**Figure 4.** a) Diversity of *meta*-C(sp$^2$)–H bond activation reactions in terms of substrates, coupling partners, catalysts, and oxidants. The color depth in each grid is proportional to the number of reactions involving each combination. b) A comparison of the theoretically accessible chemical space and realized instances in the case of the HTE and the m-CHA datasets. The numerical values provided in parentheses for each variable (molecule) are the count of distinct options observed in the dataset for that particular variable. c) Multi-Dimensional Scaling (MDS) projection of the reaction datasets.

To examine the structural diversity between our m-CHA dataset and a typical HTE dataset, we have used the multi-dimensional scaling (MDS) technique.[45] The MDS plots, as provided in Figure 4c, reveal a higher structural diversity in the m-CHA dataset as compared to that in the HTE datasets considered here. The projection of these datasets on a common 2D space indicates a more dispersed and structurally diverse distribution, mostly arising from the key constituents such as the substrates and coupling partners, in the m-CHA dataset. The HTE datasets comprise of noticeable clusters, suggestive of a relatively more homogeneous distribution of samples, whereas no such clusters are discernible the m-CHA dataset. Such diversity in the molecular structures of its participants makes the m-CHA dataset a better representative of real-world datasets. These inherent aspects of our dataset could have ramifications to the ML model performance viz-à-viz those achievable from the often-used HTE datasets (*vide infra*).

**Pre-training approach:** Training a chemical language model (CLM) requires a large amount of data, which could be time consuming and challenging due to high demands on computational resources. To address this, we initially set out to create a chemical library with relatively lesser number of molecules to pre-train a CLM. We have developed a novel substructure-based pre-training strategy, termed as SSP. First, the key substructures present in the target reaction dataset are identified by fragmenting the molecules of interest to determine unique substructures. These substructures are then mined out from the PubChem library containing in excess of 110 M commercially available molecules.[46] The idea is to generate a focused library of molecules resembling each substructure found in our target compounds of interest. This approach is likely to provide an improved representation of molecules that matter to the target specific downstream tasks. We have created three distinct SSP models, that differ in the substructures in them as sampled from the PubChem library, to pre-train the CLM (Figure 5a). These pre-training CLMs denoted as SSP1, SSP2, and SSP3 respectively consist of ~0.11 M, 0.80 M, and 6.81 M molecules represented using the corresponding SMILES. The ULMFiT model is separately pre-trained on each of these SSP datasets and the knowledge acquired is transferred to the target regressor for the desired yield prediction task.

Another important aspect that we wish to emphasize relates to the efficiency of the pre-training exercise. The recommended performance improvement measures in training LM models are to use bigger data and/or data augmentation techniques.[47] Here, we use data augmentation with a value four in the CLM pre-training, implying that the SMILES enumeration of each molecule is done using four different starting atoms. The training times on a standard hardware setting (one NVIDIA A100 gpu, 80 GB) are found to be 2.5 (SSP1), 20 (SSP2), 44 (SSP3), and 23 (ChEMBL) hrs. These are indicative of time-consuming pre-training phase when one has to deal with large molecular datasets. It also calls for customized pre-training strategies suitable for a given target task using smaller sized data as opposed to employing a conventional large dataset.

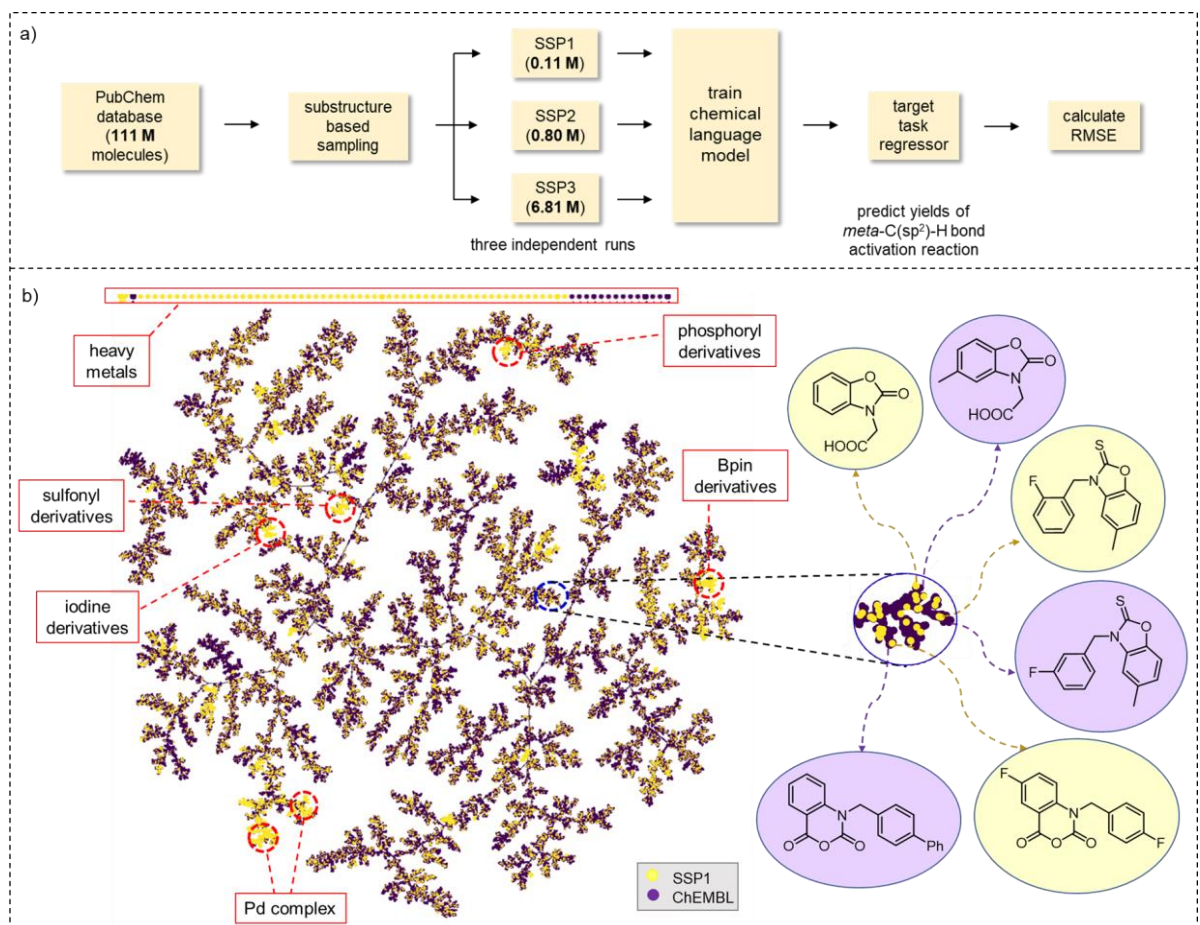

**Figure 5.** a) The ULMFiT pre-training workflow. b) The TMAP (tree map) visualization of the chemical space spanned by CHEMBL (purple) and SSP1 (yellow) datasets. The significant non-overlapping regions between two datasets are shown using red color dotted circles.

It would also be of interest to compare the chemical diversity in a relative smaller sized SSP1 dataset with those in the large ChEMBL.[48] To compare our smallest SSP1 dataset (0.11 M) with the CHEMBL (1.4 M), we employed a tree map visualization (TMAP) (Figure 5b)[49] that uses a min-hash algorithm to encode molecular SMILES and maps the chemical space. Magnifying most of the branches of the TMAP plot (indicated by a dotted blue color circle) reveals that a significant portion of the SSP1 dataset is similar to some of the closely connected clusters found within the ChEMBL dataset. To illustrate the structural similarities between the candidates in the SSP1 and ChEMBL datasets, a representative group of molecules bearing certain common substructures are shown expanded to the right side of the figure. Another interesting aspect is that a few non-overlapping regions are unique to the SSP1 dataset. These can be found in the red dotted circles as well as in the top rectangular box, latter of them belongs to molecules bearing heavy atoms and substructures missing in the ChEMBL family.[50] Therefore, we infer that a robust CLM can capture more specific chemical information from a much smaller SSP1 dataset with a much shorter pre-training time as compared to that from the large ChEMBL dataset.

**Direct regression (DR):** First, we have evaluated the influence of the size of the pre-training dataset on the ULMFiT yield prediction performance. Table 1 summarizes the performance of various pre-trained ULMFiT regressors on the m-CHA dataset.[51] It can be noticed that the model without pre-training returns inferior performance as compared to all other models using pre-training, indicating the importance of pre-training in enhancing the yield prediction accuracy. The ULMFiT regressors pre-trained on the SSP1, SSP2, SSP3, and CHEMBL show comparable performances with the test RMSEs of 10.51±0.19, 10.96±0.17, 10.89±0.19, and 10.54±0.19 respectively.[52] This observation implies that the LM could efficiently learn from the smaller pre-training datasets, which is a valuable result of high practical utility even in situations with limited training data. The ULMFiT-SSP1 regressor, with just about 8.5% pre-training samples as compared to the ChEMBL, can be considered of equivalent quality. It is also important to consider the training time with SSP1 is 2.5 hrs as opposed to 23 hrs in the case of ChEMBL.

**Table 1.** A Comparison of Performance (in terms of RMSE in % yield) of the ULMFiT Model for Yield Predictions on the m-CHA Dataset with Different Pre-training Sizes

| pre-training | Size (M) | DR | | CFR-major | | CFR-minor | |
|---|---|---|---|---|---|---|---|
| | | train | test | train | test | train | test |
| CHEMBL | 1.40 | 7.06±0.19 | 10.54±0.19 | 5.70±0.09 | 8.57±0.10 | 4.17±0.02 | 6.68±0.31 |
| **SSP1** | 0.11 | 6.81±0.12 | 10.51±0.19 | 6.04±0.17 | 8.40±0.12 | 4.21±0.08 | 6.48±0.29 |
| SSP2 | 0.80 | 6.72±0.03 | 10.96±0.17 | 5.14±0.03 | 8.54±0.11 | 4.00±0.02 | 6.85±0.32 |
| SSP3 | 6.81 | 6.33±0.04 | 10.89±0.19 | 5.03±0.02 | 8.54±0.12 | 4.01±0.02 | 6.72±0.31 |
| none | - | 7.81±0.42 | 11.28±0.16 | 7.27±0.25 | 8.97±0.26 | 5.34±0.16 | 7.53±0.31 |

A comparison of the ULMFiT-SSP1 performance with the other commonly used LMs within the realm of chemical reactivity is also considered here. The transformer based LMs trained on the concatenated reactant SMILES led to a relatively inferior performance (RMSEs of Yield-BERT and Yield-BERT-DA respectively are 11.36±0.13 and 11.48±0.13). Similarly, the performance of a molecular fingerprint-based transformer such as the FP-BERT (uses fingerprint based BERT encoded features) as well as the graph-based neural networks (Graph-RXN and MPNN) turned out to be slightly inferior to the ULMFiT.[53] Although the ULMFiT-SSP1 model gave a good RMSE of 10.51±0.19, implying ~70% of the predictions are within 10 units of the actual experimentally known yield, we wanted to examine whether the ability of the model could be improved. There are about 30% predictions with differences larger than 10 units in %yield, indicating certain latent challenges in the generalizability of the DR approach. It is quite possible that such performance issues might stem from the distribution characteristics in our dataset as described earlier in this manuscript, such as the class imbalance and sparsity. In light of these, we designed a novel model termed as CFR, which classifies the data prior to applying a regression model, as described below.

**Classification followed by regression (CFR):** The unevenly distributed ground truth yield values as found in the reported wet-lab experiments in the m-CHA dataset presents a case of class imbalance,[10] with a dominant share of the data belonging to the high %yield region, leaving only a very few samples with lower yields. To tackle this issue, a classifier is developed to stratify the dataset into multiple classes based on their output distributions. First, the Bayes error estimator (BER)[54] is employed to determine the optimal class boundary for the classifier that assigns discrete labels to each reaction sample as high or low. Subsequently, separate regression models are built for the major and minor classes. This integrated approach, shown in Figure 6a, might help in making yield predictions more robust for both these classes.

The choice of the class boundaries in our CFR model is made on the basis of the natural distribution of the yield as seen in the m-CHA reactions. From a statistical perspective, the class boundaries for a binary classification could be placed at $\mu$, $(\mu+\sigma)$, or $(\mu-\sigma)$, where $\mu$ and $\sigma$ are the mean and standard deviation, respectively. In our dataset, $\mu$ of the % yield is 66.10 and $\sigma$ is 12.82. The BER analysis reveals that the maximum achievable classification accuracy is 96.9% with $(\mu–\sigma)$ as the class boundary.[55] The reactions with experimentally reported %yield, ranging from 0 to 53%, are therefore categorized as the CFR-minor class (leading to 155 reactions), and those higher than this threshold form the CFR-major class (711 reactions).

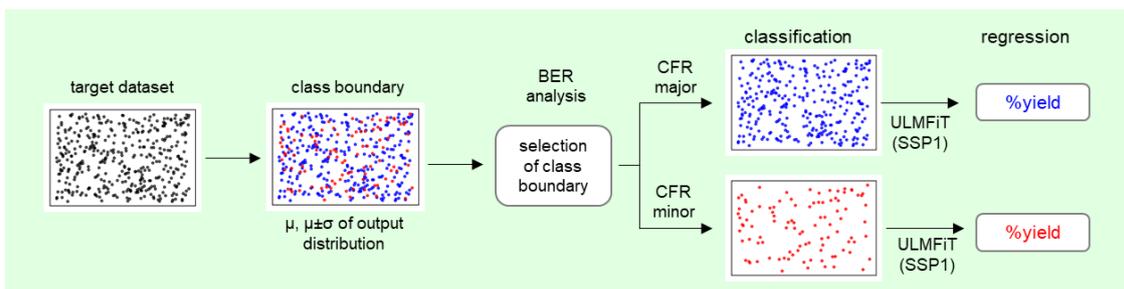
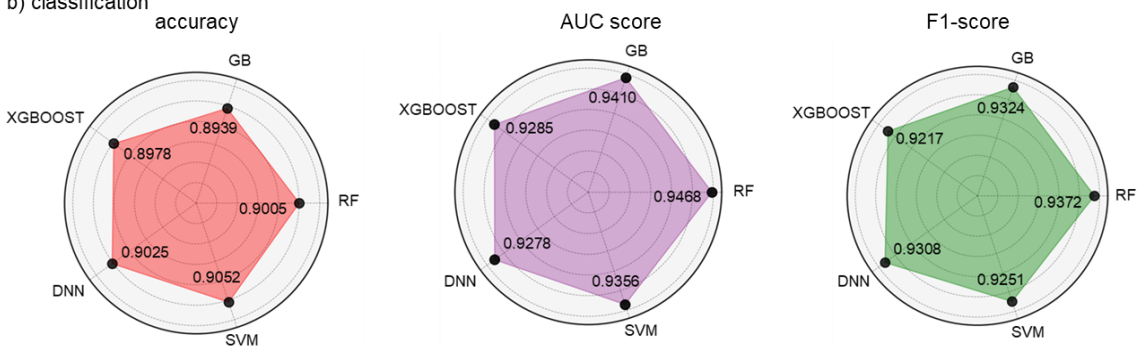
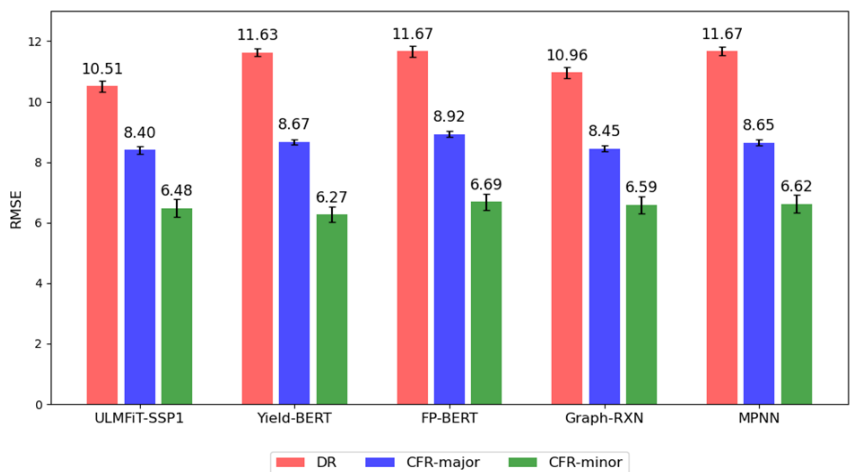

**Figure 6.** a) A general overview of our classification followed by regression (CFR) approach. b) Performance comparison of different classification models using the average F1 score, AUC-score, and accuracy for the test sets. c) Performance comparison of different yield prediction models using the test RMSEs (expressed in %yield) on the m-CHA reaction dataset as obtained through the DR approach, CFR-major, and CFR-minor classes. The error bars denote the corresponding standard error in each case.

After identifying the optimal class boundary for the CFR implementation, we shifted our attention toward developing a classification model capable of distinguishing the major and

>19

minor class samples in our dataset. Five different classifiers based on random forest (RF), gradient boosting (GB), extreme gradient boosting (XGBOOST), support vector machine (SVM), and deep neural network (DNN) are considered.[56] Here, the 400 dimensional encoder output as obtained from the ULMFiT-SSP1 model, serves as the input to these classification models. The performance of the classifier is evaluated using standard metrics such as accuracy, F1-score, and AUC-score (area under the receiver operating characteristics curve).

Since our target dataset is skewed and class imbalanced, we have used the SMOTE (synthetic minority oversampling)[57] technique to generate additional synthetic samples for the underrepresented CFR-minor class for training the classifiers.[58] As can be gleaned from Figure 6b, the test accuracy of 0.9025, F1-score of 0.9308, and AUC score of 0.9278 are obtained for the DNN classifier. All other classifier models such as the RF, GB, XGBOOST, and SVM also exhibited better performance with the inclusion of SMOTE samples. Since the DNN classifier is found to be a superior classifier as indicated by all the three performance metrics, we have developed a robust DNN-based classification model to categorize reactions into major and minor classes. Aided by a good quality classifier, we have subsequently focused on developing separate regressors for the CFR-major and -minor classes.

For the regression tasks, we have evaluated the performance of multiple models such as the ULMFiT-SSP1, transformer, and graph-based regressors for the CFR-major and CFR-minor classes, results of which are provided in Figure 6c. The ULMFiT-SSP1 regressor provides an impressively good test RMSEs of 8.40±0.12 and 6.48±0.19 respectively for the CFR-major and CFR-minor classes. This is a significant improvement over that obtained from the DR model with an RMSE of 10.51±0.19 in %yield. In addition, we find that the ULMFiT regressor with other larger pre-training datasets such as SSP2, SSP3, and ChEMBL offer comparable performances to that of ULMFiT-SSP1. However, the corresponding ULMFiT model without pre-training is found to be notably inferior to those with pre-training (Table 1).

A similar performance of the ChEMBL based pre-training as compared to the SSP models, despite the former lacking compounds containing heavy elements, could possibly stem from the considerable similarity (~98%) in the share of key elements in the training data.[57]

The most important observation is that the ULMFiT-SSP1 model outperforms, both in the major and minor class regression tasks, over other models such as the transformers (Yield-BERT and Yield-BERT-DA) and graph-based models (MPNN and Graph-RXN). We have also evaluated the confidence intervals (CIs) to quantify the uncertainty estimates by providing a range within which the true population parameter is likely to fall. For the CFR-major class, the RMSE is estimated to fall between 8.15 and 8.64 with 95% confidence, meaning that the true RMSE is likely to be within this range. In the case of the CFR-minor class, the corresponding window of the RMSE is between 5.87 and 6.29, with a CI of 95%. This indicates the robustness of the CFR model in predicting the yields of reactions belonging to both major and minor classes. The key take home at this juncture is the superior predictive efficiency of our CFR model over the DR model (Figure 6c).

A direct comparison of the predictive capabilities of the DR and CFR models can be drawn from Figure 7a, wherein $\Delta$yield that captures the difference between the experimentally reported ground truth yield and that predicted by our models, is provided. A detailed analysis reveals that about 5% of samples exhibit $\Delta$yield >20 units in DR. In contrast, with the CFR-major class only 2% predictions exceed this threshold of 20 units while the CFR-minor is even better with as little as 1% samples going beyond this boundary. These are clear indicators of the superior yield predictions offered by the CFR model. To examine this interesting result further, we have first identified the test samples that exhibit $\Delta$yield >20 units in the DR model. The corresponding $\Delta$yield obtained from the CFR predictions is then compared (Figure 7b). It is readily discernible that a large number of predictions from the DR are above the threshold error of 20 units (shown using a horizontal dashed line). On the other hand, the error for majority of the CFR predictions is well below this threshold. This is quite assuring of the efficacy of our CFR model towards enhancing the quality of predictions for a complex reaction such as the *meta*-C(sp$^2$)–H bond activation.

Apart from the impressive performance of the CFR model over the DR, the robustness and generalizability of the former could be seen across different control experiments that we have considered. These runs such as a) randomization of the classification labels to evaluate the learning efficacy,[59] b) learning curve analysis to assess the impact of training size,[60] and c) performance on new holdout test sets to evaluate consistency in model performance with our dataset,[61] are all found to be convincing. On the basis of all these control runs and the evaluations, we propose that the CFR model is more suitable for yield predictions for transition metal catalyzed *meta*-C(sp$^2$)–H bond activation reactions. The same approach should hold good for other reactions as well, as our approach directly addresses the inherent distribution issues often found in chemical reaction datasets.

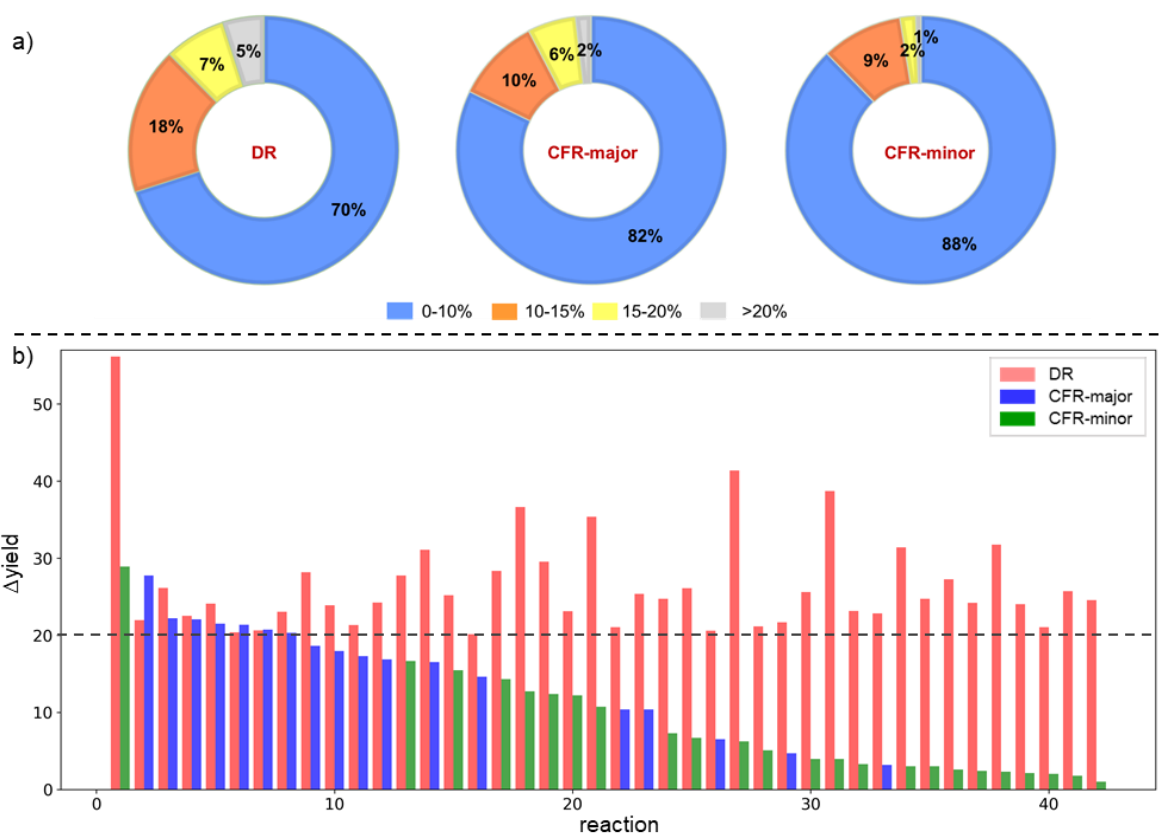

**Figure 7.** a) Pie chart of Δyield (difference between the experimentally reported and predicted %yield) for test samples obtained using the ULMFIT-SSP1 model for DR, CFR-major, and CFR-minor cases. b) A bar plot comparing Δyield of test samples as obtained from the DR (red color) and CFR-major (blue) and CFR-minor (green) models.

**Benchmarking studies of ML models on the *meta*-C(sp$^2$)–H bond activation dataset:** Here, we undertake an important comparison of the model performances obtained on the m-CHA dataset with other HTE datasets. The BER analysis, as described in the previous section, conveys that the maximum achievable classification accuracy (79% for the Buchwald-Hartwig (BH) and 82% for the Suzuki (SC) datasets) would be when the corresponding mean of % yield is chosen as the class boundary. The application of the CFR model to these commonly used HTE datasets shows good classification accuracies of 0.9189 (BH) and 0.8928 (SC). More significant aspect is the substantial boost in regression performance obtained for both the CFR-major and CFR-minor classes for these HTE datasets (Table 2). It should be considered that all previous studies on these HTE datasets could not give performances as good as our CFR model.[25,26,38,45,62] For instance, the previously reported best RMSEs for the BH dataset was 4.36±0.03, which is inferior to 3.68 (CFR-major), where most of the reactions belong to. However, an RMSE of 5.36 for the CFR-minor is not as good as previous models. Similarly, the test RMSE of 9.23±0.13 in the case of the SC dataset is surpassed by our CFR model (8.05 (major) and 9.04 (minor)).[81d] Thus, our CFR model can be considered the state-of-the-art for the key three datasets considered in this study.

This curious observation regarding the model performance may be attributed to the inherent bias in our dataset as shown in Figure 4. The HTE datasets present a relatively more homogeneous chemical space encompassing all reactions between its reacting partners and carries very low experimental or reporting biases. Unlike in our manually curated m-CHA dataset, HTE documents even the low yielding reactions besides the use of standardized experimental conditions allowing for very little variance in measurements. These characteristics render the HTE datasets better suited for case studies for statistical learning than the m-CHA dataset, albeit at the cost of exploring a much narrower chemical space due to much less diversity in its reacting partners (Figure 4c). In other words, ML model evaluations based only on the HTE datasets should be considered with caution, as they are less likely to perform well in real-world situations such as with the m-CHA dataset. Hence, a distribution aware CFR model as proposed in this work would be a better alternative for chemical reaction outcome predictions.

**Table 2.** A Comparison of the ULMFiT-SSP1 Performance Across Different Datasets Reported as the Averaged Over 20 Independent Runs

| dataset | reaction | DR | CFR-major | CFR-minor |
|---------|----------|-----|-----------|-----------|
| BH | Buchwald-Hartwig coupling | 5.62±0.08 | 3.68±0.14 | 5.36±0.08 |
| SC | Suzuki coupling | 10.06±0.19 | 8.05±0.13 | 9.04±0.08 |
| m-CHA | *meta*-C(sp$^2$)−H activation | 10.51±0.19 | 8.40±0.12 | 6.48±0.29 |
| NiCOlit | nickel catalyzed C−O coupling | 22.15±0.62 | 17.34±0.50 | 5.81±0.24 |
| ELN | Buchwald-Hartwig coupling | 22.58±0.69 | 19.29±0.62 | 2.46±0.11 |
| AH | asymmetric hydrogenation | 8.48±0.35 | 2.70±0.09 | 12.16±0.89 |
| USPTO | combination of different reaction classes | 0.21±0.01 | 0.20±0.01 | 0.02±0.00 |

In going beyond the three important reaction datasets thus far considered in our study, we have evaluated the performance of our ULMFiT model across four additional datasets as frequently found in literature. The results provided in Table 2 indicate that similar to the superior performance of the CFR model obtained with the HTE datasets (BH and SC), it also does better for many other real-world datasets such as NiCOlit (RMSE 22.80)[38a] and ELN (25.27).[38b] A comparable performance is obtained on the USPTO as well.[63]

However, an exception is observed with a highly imbalanced AH dataset, where the DR model offers a slightly better performance than the CFR.[82] In light of this, we became interested in evaluating the limitations of our CFR model as well as to make our recommendations clear as to when would it be better to deploy the DR model. In other words, for what kind of data specific situations one should prefer DR to CFR. The analysis of the output distribution and the model performance across different datasets revealed that the skewness ($\gamma$) could serve as an early indicator in making an informed choice between the DR and CFR models. It is noted that for datasets with higher asymmetry, as indicated by the $\gamma$ values lesser than -1 and greater than +1, the CFR model is unlikely to outperform the corresponding DR model.[64] Hence, we suggest the use of the CFR model when the $\gamma$ of the output distribution is in the range [-1,1], which is a likely situation in most datasets in common use today.

**Conclusions**

In keeping with the contemporary interest in utilizing machine learning (ML) for chemical applications, we developed a novel approach for yield prediction suitable for sparse and imbalanced data distributions as often found in chemical reaction development. First, we contribute a manually curated reaction dataset comprising of more than 800 synthetically important *meta*-C(sp$^2$)−H activation reactions (m-CHA) of high contemporary interest. Unlike high-throughput experimentation (HTE) datasets, the m-CHA dataset is notably sparse and spans a wider chemical space, suggestive of experimental selection bias toward certain type of catalyst/substrate during reaction development. Direct deployment of standard deep learning built on chemical language models for yield prediction on the m-CHA reactions, with and without pre-training on large chemical databases, generally led to lower performance. Unlike the prevailing pre-training practices wherein one would use large library of unlabeled molecules directly from ChemBL database, we propose a novel substructure based pre-training strategy, where a new library of 0.11 M molecules of relevance to the candidate molecules in the target task are first mined out from the large PubChem database to give SSP1, which is then employed for pre-training of the ULMFiT (Universal Language Model Fine Tuning) model. Consequently, our ULMFiT model effectively learns data-specific chemical language, with a training time of just 2.5 hrs with our SSP1 pre-training dataset with only about 8.5% of the size of ChEMBL. This approach assures a time- and resource-efficient alternative to pre-training using bigger datasets. Notably, the ULMFiT model pre-trained on both ChEMBL and SSP1, provides comparable performances. This indicates that a focused, smaller dataset like SSP1 can capture sufficient chemical information for effective pre-training.

Since the output distribution in our m-CHA dataset is skewed toward the higher values, we propose a novel model, denoted as CFR (classification followed by regression) that does a classification prior to regression. A given reaction is first identified as belonging to a 'major' or a 'minor' class with respect to a statistically meaningful class boundary that uses the mean ($\mu$) and standard deviation ($\sigma$) of the yield values. The classified samples are subsequently sent to either of the two independent regressors, CFR-major or CFR-minor, built on a fine-tuned chemical language model based on the ULMFiT architecture to predict the yield of the reaction. The test RMSE of the ULMFiT-SSP1 regressor is found to be 8.40±0.12 for the CFR-major class and 6.48±0.29 for the CFR-minor class, significantly outperforming a direct regression model (DR), devoid of prior classification. The CFR approach improved prediction quality, with only 2% of samples in the CFR-major class and 1% in the CFR-minor class exhibiting Δyield>20 units as opposed to 5% predictions above this threshold for the DR model. The generalizability of our CFR model remains impressive over other widely used datasets such as Buchwald-Hartwig coupling, Suzuki coupling, nickel catalyzed C−O coupling, and USPTO, as it could provide the state-of-the-art test accuracies. However, it outperforms the DR model when the skewness in the output distribution falls within the range of [-1, 1]. Thus, we could develop a robust ML model for yield prediction which can be deployed on a diverse range of chemical reaction datasets, which could be useful in reaction development and for exploration of the untested reaction space.

**ASSOCIATED CONTENT**

**Corresponding Author**

**Raghavan B. Sunoj**: Department of Chemistry and Centre for Machine Intelligence and Data Science, Indian Institute of Technology Bombay, Mumbai, Maharashtra, 400076, India; orcid.org/0000-0002-6484-2878; Email- sunoj@chem.iitb.ac.in

**Authors**

**Supratim Ghosh**: Department of Chemistry, Indian Institute of Technology Bombay, Powai, Mumbai 400076.

**Nupur Jain**: Department of Chemistry, Indian Institute of Technology Bombay, Powai, Mumbai 400076.

**Author contribution**: S.G. and N.J. contributed equally.

**Acknowledgements**

We are thankful to Institution of Eminence (IoE) Data and Information Science computing facility for generous computational resources. S.G. and N.J respectively acknowledge the council of scientific and industrial research for a senior research fellowship and the prime ministers research fellowship. We are grateful to Preethi Jyothi (Department of Computer Science and Engineering, IIT Bombay) for some valuable discussions during the course of this work.

**Notes**

The authors declare no competing financial interest.

**Code and data availability**

The datasets and source code utilized for training the models are accessible on GitHub: https://github.com/Nupurjain2788/m-CHA-CFR-yield-prediction.git

52. The test RMSE in the case of the y-scrambled dataset is found to be as high as 15.08±0.12, reflecting a significant decline in performance. This observation indicates that the model is able to effectively learn molecular features of the input reaction to be able to predict the yield of the reaction.

53. a) The FPBERT, GraphRXN, and MPNN models give an average test RMSE of 11.67±0.18, 10.96±0.19, and 11.67±0.14 respectively for 20 independent runs. b) The implication of concatenated reactant and product SMILES (reactant SMILES >>product SMILES) is evaluated using the ULMFiT-SSP1 model for a representative case to learn that it did not improve the performance.

54. Noshad, M.; Xu, L.; Hero, A. Learning to Benchmark: Determining Best Achievable Misclassification Error From Training Data. *arXiv*, **2019**.

55. a) When the class boundary is set at $\mu$, the affordable classification accuracy according to the BER estimator is 78.5 %, while for ($\mu+\sigma$) it is 84.9 %. b) The accuracies for tertiary and quaternary classifications are <70%.

56. a) For the classification models, we used 80% of the dataset for training. During training, a grid hyperparameter search method is used for identifying the optimal hyperparameters. The model was then evaluated on the remaining 20% of the dataset, as the test set. b) The ULMFiT classifier pre-trained on the SSP1 showed an average accuracy, F1-score, and AUC-score of 0.8247, 0.8117, and 0.8416, respectively, on the test set. c) We note that the training of ULMFIT-SSP1 classifier with different degree of SMILES augmentation did not improve the accuracy of the model. d) The inclusion of a weighted random sampler in the minority class data could not improve the performance of the ULMFiT-SSP1 model. e) The performance of all the five classification models (RF, GB, XGBOOST, SVM, and DNN) are comparable to the ULMFiT-SSP1, with an average test accuracy >0.81, F1 score >0.89, and AUC value >0.73.

57. a) Chawla, N. V.; Bowyer, K. W.; Hall, L. O.; Kegelmeyer, W. P. SMOTE: Synthetic Minority Over-sampling Technique. *J. Artif. Intell.* **2002**, *16*, 321. b) Demidova, L.; Klyueva, I. SVM Classification: Optimization with the SMOTE Algorithm for the Class Imbalance Problem, *6th Mediterranean Conference on Embedded Computing* **2017**. c) Douzas, G.; Bacao, F. Geometric SMOTE: Effective Oversampling for Imbalanced Learning Through a Geometric Extension of SMOTE. *arXiv*, **2017**.

58. a) The SMOTE technique could provide 556 additional samples in the CFR-minor class, totaling to 1422 instances. It should be noted that the synthetic data is used only in the model training, while the test set contains only real samples. For previous instances of using SMOTE technique in molecular machine learning applications can be found in b) Ying, D.; Hua, P.; Hao, M. Research and Application of SMOTE-Based Method with XGBoost Regression Prediction. *2023 IEEE International Conference on Image Processing and Computer Applications (ICIPCA)* **2023**, 1737. c) Mahmud, S. M. H.; Chen, W.; Jahan, H.; Liu, Y.; Sujan, N. I.; Ahmed, S. IDTi-CSsmoteB: Identification of Drug–Target Interaction Based on Drug Chemical Structure and Protein Sequence Using XGBoost with over-Sampling Technique SMOTE. *IEEE Access* **2019**, *7*, 48699.

59. a) A CFR model is trained by using randomizing the classification labels in such a way that each sample is largely mapped to an incorrect label. A notably poorer performance (classification accuracy 0.7417, regression RMSE of 11.44±0.14 and 12.66±0.46 respectively for the CFR-major and -minor classes in %yield) with the label randomization compared to when the true labels were used suggests that the LM is effectively learning the classification into major and minor groups and that the regression works better for the individual classes. b) We have also considered a CFR model by gradually increasing the misclassified labels from 10% to 100%. A decrease in the CFR model performance with an increase in misclassified samples could be seen. This analysis indicates that the CFR model truly learns the features provided from the input featurization.

60. The CFR model is evaluated by progressively increasing the training size from 40% to 80%. The analysis demonstrates that the ULMFiT-SSP1 model exhibits minimal differences in the train and test performances across these training sizes, suggesting that no significant overfitting or underfitting issues prevail.

61. We have created 10 new holdout test sets of 100 randomly chosen samples from among the full set of 866 reactions. The newly trained CFR model with 766 reactions exhibited an impressively good classification accuracy of 0.8410 (average over 10 runs), an F1 score of 0.8989, and an AUC score of 0.7891. Another interesting aspect is that the test RMSEs of 8.57±0.06 (CFR-major) and 6.90±0.12 (CFR-minor) are quite comparable to the performances obtained with the full dataset.

62. a) Han, J.; Kwon, Y.; Choi, Y.-S.; Kang, S. Improving Chemical Reaction Yield Prediction Using Pre-Trained Graph Neural Networks. *J. Cheminform.* **2024**, 16, 25. b) Zhao, W.; Li, Y. Predicting the Yield of Pd-catalyzed Buchwald–Hartwig Amination Using Machine Learning with Extended Molecular Fingerprints and Selected Physical Parameters. *ChemistrySelect* **2024**, *9*, 33. c) Hu, W.; Liu, B.; Gomes, J.; Zitnik, M.; Liang, P.; Pande, V.; Leskovec, J. Strategies for Pre-Training Graph Neural Networks. *arXiv*, **2019**. d) Chen, J.; Guo, K.; Liu, Z.; Isayev, O.; Zhang, X. Uncertainty-Aware Yield Prediction with Multimodal Molecular Features. *Proc. Conf. AAAI Artif. Intell.* **2024**, *38*, 8274.

63. Yin, X.; Hsieh, C.-Y.; Wang, X.; Wu, Z.; Ye, Q.; Bao, H.; Deng, Y.; Chen, H.; Luo, P.; Liu, H.; Hou, T.; Yao, X. Enhancing Generic Reaction Yield Prediction through Reaction Condition-Based Contrastive Learning. *Research* **2024**, *7*, 0292.

64. The natural skewness in the output distribution observed in the case of BH, SC, m-CHA, AH, NiCOlit, and ELN datasets are 0.51, 0.44, -0.25, -2.79, -0.44, and 0.16 respectively.